\documentclass[twocolumn,prc,preprintnumbers,superscriptaddress,showpacs,amsmath,amssymb,floatfix]{revtex4}
\def\e+e-{$e^+e^-$}

\def\eeprotonpi0{$^1$H$(e,e^{\prime}p)\pi^0$ }

\def\qqbar{$q\overline q$ }
\def\ssbar{$s\overline s$ }
\def\ddbar{$d\overline d$ }
\def\uubar{$u\overline u$ }

\def\qsq{$Q^2$}

\def\kpluslam{$\Lambda K^+$}
\def\neutronpiplus{$n\pi^+$}
\def\ppi0{$p\pi^0$}

\def\costh{$\cos\theta^*_m$ }

\def\threep0{$^3P_0$ }
\def\threes1{$^3S_1$ }

\usepackage[dvips]{graphicx}
\usepackage{latexsym}
\usepackage{footnote}
\usepackage{colordvi}
\usepackage{fancyhdr}
\usepackage{amssymb}
\usepackage{url}
\usepackage{longtable}
\usepackage{dcolumn}
\usepackage{rotating}
\usepackage{ae}
\usepackage{aecompl}

\begin{document}
\vskip 2.cm

\title{Strangeness Suppression in \qqbar Creation Observed in
Exclusive Reactions}
\vskip 1.cm

\newcommand*{\ANL}{Argonne National Laboratory, Argonne, Illinois 60439, USA}
\newcommand*{\ANLindex}{1}
\affiliation{\ANL}
\newcommand*{\CANISIUS}{Canisius College, Buffalo, New York, USA}
\newcommand*{\CANISIUSindex}{2}
\affiliation{\CANISIUS}
\newcommand*{\CSUDH}{California State University, Dominguez Hills, Carson, California 90747, USA}
\affiliation{\CSUDH}
\newcommand*{\CMU}{Carnegie Mellon University, Pittsburgh, Pennsylvania 15213, USA}
\newcommand*{\CMUindex}{3}
\affiliation{\CMU}
\newcommand*{\CUA}{Catholic University of America, Washington, D.C. 20064, USA}
\newcommand*{\CUAindex}{4}
\affiliation{\CUA}
\newcommand*{\SACLAY}{CEA, Centre de Saclay, Irfu/Service de Physique Nucl\'eaire, 91191 Gif-sur-Yvette, France}
\newcommand*{\SACLAYindex}{5}
\affiliation{\SACLAY}
\newcommand*{\CNU}{Christopher Newport University, Newport News, Virginia 23606, USA}
\newcommand*{\CNUindex}{6}
\affiliation{\CNU}
\newcommand*{\UCONN}{University of Connecticut, Storrs, Connecticut 06269, USA}
\newcommand*{\UCONNindex}{7}
\affiliation{\UCONN}
\newcommand*{\FU}{Fairfield University, Fairfield Connecticut 06824, USA}
\newcommand*{\FUindex}{8}
\affiliation{\FU}
\newcommand*{\FIU}{Florida International University, Miami, Florida 33199, USA}
\newcommand*{\FIUindex}{9}
\affiliation{\FIU}
\newcommand*{\FSU}{Florida State University, Tallahassee, Florida 32306, USA}
\newcommand*{\FSUindex}{10}
\affiliation{\FSU}
\newcommand*{\GWUI}{The George Washington University, Washington, D.C. 20052, USA}
\newcommand*{\GWUIindex}{11}
\affiliation{\GWUI}
\newcommand*{\ISU}{Idaho State University, Pocatello, Idaho 83209, USA}
\newcommand*{\ISUindex}{12}
\affiliation{\ISU}
\newcommand*{\INFNFE}{INFN, Sezione di Ferrara, 44100 Ferrara, Italy}
\newcommand*{\INFNFEindex}{13}
\affiliation{\INFNFE}
\newcommand*{\INFNFR}{INFN, Laboratori Nazionali di Frascati, 00044 Frascati, Italy}
\newcommand*{\INFNFRindex}{14}
\affiliation{\INFNFR}
\newcommand*{\INFNGE}{INFN, Sezione di Genova, 16146 Genova, Italy}
\newcommand*{\INFNGEindex}{15}
\affiliation{\INFNGE}
\newcommand*{\INFNRO}{INFN, Sezione di Roma Tor Vergata, 00133 Rome, Italy}
\newcommand*{\INFNROindex}{16}
\affiliation{\INFNRO}
\newcommand*{\ORSAY}{Institut de Physique Nucl\'eaire ORSAY, Orsay, France}
\newcommand*{\ORSAYindex}{17}
\affiliation{\ORSAY}
\newcommand*{\ITEP}{Institute of Theoretical and Experimental Physics, Moscow, 117259, Russia}
\newcommand*{\ITEPindex}{18}
\affiliation{\ITEP}
\newcommand*{\JMU}{James Madison University, Harrisonburg, Virginia 22807, USA}
\newcommand*{\JMUindex}{19}
\affiliation{\JMU}
\newcommand*{\KNU}{Kyungpook National University, Daegu 702-701, Republic of Korea}
\newcommand*{\KNUindex}{20}
\affiliation{\KNU}
\newcommand*{\LPSC}{LPSC, Universit\'e Grenoble-Alpes, CNRS/IN2P3, Grenoble, France}
\affiliation{\LPSC}
\newcommand*{\UNH}{University of New Hampshire, Durham, New Hampshire 03824-3568, USA}
\newcommand*{\UNHindex}{21}
\affiliation{\UNH}
\newcommand*{\NSU}{Norfolk State University, Norfolk, Virginia 23504, USA}
\newcommand*{\NSUindex}{22}
\affiliation{\NSU}
\newcommand*{\OHIOU}{Ohio University, Athens, Ohio  45701, USA}
\newcommand*{\OHIOUindex}{23}
\affiliation{\OHIOU}
\newcommand*{\ODU}{Old Dominion University, Norfolk, Virginia 23529, USA}
\newcommand*{\ODUindex}{24}
\affiliation{\ODU}
\newcommand*{\RPI}{Rensselaer Polytechnic Institute, Troy, New York 12180-3590, USA}
\newcommand*{\RPIindex}{25}
\affiliation{\RPI}
\newcommand*{\URICH}{University of Richmond, Richmond, Virginia 23173, USA}
\affiliation{\URICH}
\newcommand*{\ROMAII}{Universita' di Roma Tor Vergata, 00133 Rome Italy}
\newcommand*{\ROMAIIindex}{26}
\affiliation{\ROMAII}
\newcommand*{\MSU}{Skobeltsyn Institute of Nuclear Physics, Lomonosov Moscow State University, 119234 Moscow, Russia}
\newcommand*{\MSUindex}{27}
\affiliation{\MSU}
\newcommand*{\SCAROLINA}{University of South Carolina, Columbia, South Carolina 29208, USA}
\newcommand*{\SCAROLINAindex}{28}
\affiliation{\SCAROLINA}
\newcommand*{\JLAB}{Thomas Jefferson National Accelerator Facility, Newport News, Virginia 23606, USA}
\newcommand*{\JLABindex}{29}
\affiliation{\JLAB}
\newcommand*{\UTFSM}{Universidad T\'{e}cnica Federico Santa Mar\'{i}a, Casilla 110-V Valpara\'{i}so, Chile}
\newcommand*{\UTFSMindex}{30}
\affiliation{\UTFSM}
\newcommand*{\EDINBURGH}{Edinburgh University, Edinburgh EH9 3JZ, United Kingdom}
\newcommand*{\EDINBURGHindex}{31}
\affiliation{\EDINBURGH}
\newcommand*{\GLASGOW}{University of Glasgow, Glasgow G12 8QQ, United Kingdom}
\newcommand*{\GLASGOWindex}{32}
\affiliation{\GLASGOW}
\newcommand*{\VIRGINIA}{University of Virginia, Charlottesville, Virginia 22901, USA}
\newcommand*{\VIRGINIAindex}{33}
\affiliation{\VIRGINIA}
\newcommand*{\WM}{College of William and Mary, Williamsburg, Virginia 23187-8795, USA}
\newcommand*{\WMindex}{34}
\affiliation{\WM}
\newcommand*{\YEREVAN}{Yerevan Physics Institute, 375036 Yerevan, Armenia}
\newcommand*{\YEREVANindex}{35}
\affiliation{\YEREVAN}

\newcommand*{\NOWUTFSM}{Universidad T\'{e}cnica Federico Santa Mar\'{i}a, Casilla 110-V Valpara\'{i}so, Chile}
\newcommand*{\NOWJLAB}{Thomas Jefferson National Accelerator Facility, Newport News, Virginia 23606, USA}
\newcommand*{\NOWODU}{Old Dominion University, Norfolk, Virginia 23529, USA}
\newcommand*{\NOWUCONN}{University of Connecticut, Storrs, Connecticut 06269, USA}
\newcommand*{\NOWGLASGOW}{University of Glasgow, Glasgow G12 8QQ, United Kingdom}
\newcommand*{\NOWORSAY}{Institut de Physique Nucl\'eaire ORSAY, Orsay, France}
\author {M.D.~Mestayer} 
\affiliation{\JLAB}
\author {K.~Park} 
\altaffiliation[Current address:]{\NOWODU}
\affiliation{\JLAB}

\author {K.P. ~Adhikari} 
\affiliation{\ODU}
\author {M.~Aghasyan} 
\affiliation{\INFNFR}
\author {S. ~Anefalos~Pereira} 
\affiliation{\INFNFR}
\author {J.~Ball} 
\affiliation{\SACLAY}
\author {M.~Battaglieri} 
\affiliation{\INFNGE}
\author {V.~Batourine} 
\affiliation{\JLAB}
\author {I.~Bedlinskiy} 
\affiliation{\ITEP}
\author {A.S.~Biselli} 
\affiliation{\FU}
\affiliation{\CMU}
% ...

\collaboration{The CLAS Collaboration}
\noaffiliation

\vskip 1.cm

\begin{abstract}
We measured the ratios of 
electroproduction cross-sections from a proton target for three
exclusive meson-baryon final states: $\Lambda K^+$, $p\pi^0$, and $n\pi^+$,
with the CLAS detector at Jefferson Lab.  
Using a simple model of quark hadronization we extract \qqbar creation probabilities
for the first time in exclusive two-body production, in which only a single \qqbar pair is
created.  We observe a sizable suppression of strange quark-antiquark pairs compared
to non-strange pairs,
similar to that seen in high-energy production.  
\end{abstract}

\maketitle


At high energies the production of hadrons is
well described by a model in which the color ``flux-tube''
is ``broken'' by a series of \qqbar pair creation events
followed by a regrouping of the quarks and anti-quarks into
color singlet hadrons.   
The modeling of the strong force as a color flux tube 
explained the linear binding potential of heavy \qqbar ``quarkonia''
states, while quark-pair creation models~\cite{micu} developed in the
1970's accounted for hadronic production and the non-observance of free quarks.

The ``Lund Model''~\cite{lund} was formulated in the 1980's to quantify the 
fragmentation of very high-momentum
quarks into ``jets'' of observed hadrons.
The \qqbar pair creation process is modeled as tunneling in a 
linear potential, resulting in a probability proportional to the
exponential of the quark mass squared divided by the flux-tube
tension of $\approx 1$ GeV/fm.
Calculations with plausible quark masses indicated that \ssbar
production is reduced by a factor of about one-third relative to that for \uubar or $d\overline d$.
This reduction factor is known as the ``strangeness suppression factor'',
and is empirically adjusted to
approximate the observed production rates of hadrons.

Strangeness suppression has been studied
by various hadron-production experiments~\cite{suppressionresults} resulting in a successful
extension of the Lund Model into, among others, the JETSET and PYTHIA event generators~\cite{PYTHIAWeb1}
which reproduce observed hadronic production rates in high energy reactions.
Typically, a strangeness suppression factor, $\lambda_s$ $\approx$
0.3 describes the data well in $e^+e^-$ collisions up to center-of-mass energies of the $Z$ boson mass~\cite{zmasssuppression} and in high-energy deep-inelastic electron proton scattering~\cite{hera}.

Although \qqbar creation is the ``kernel'' of the process which transforms
quarks into observable hadrons, it is not well-understood.  We designed our study
to extract the flavor-dependence of \qqbar creation
in a new kinematic region: the two-body exclusive limit in which a single \qqbar pair is created,
there are no decay chains to model and for which we can do an explicit phase-space correction.

In pseudoscalar-meson 
electroproduction, a beam of electrons is incident upon a proton target, 
producing a final state consisting of the scattered electron
and the outgoing baryon and pseudo-scalar meson.   
After integrating over the azimuthal angle 
of the scattered electron, the cross section can be expressed 
in terms of the variables $Q^2$, $W$, $\theta_m^*$, and $\phi$, where 
$q^2=-Q^2$ is the squared four-momentum of the virtual photon,
$W$ is the total hadronic energy in the center-of-mass frame, 
$\theta^*_{m}$ is the meson angle in the $\gamma^*p$ center-of-mass system and 
$\phi$ is the azimuthal angle of the reaction plane with respect to the electron scattering plane:

The differential cross-section can be expressed as:
\begin{eqnarray}
\frac{d\sigma}{dQ^2 dW d\Omega^*_m} = \Gamma_v \Big( \sigma_T + \epsilon \sigma_L + \epsilon \sigma_{TT} \cos 2\phi \nonumber\\
+ \sqrt{\epsilon(\epsilon+1)}\sigma_{LT} \cos \phi \Big)~,
\end{eqnarray}
\noindent
where $\Gamma_v$ is the flux of virtual photons,
$\epsilon$ is the polarization parameter, and the
four structure functions, $\sigma_T$, $\sigma_L$, $\sigma_{TT}$ and $\sigma_{LT}$ are the
transverse and longitudinal response functions and the two interference terms, respectively.
This formalism is explained in more detail in a previous CLAS collaboration paper~\cite{5str}.


Our study was part of a larger program to measure electroproduction
of hadrons from a proton target.  The electron beam 
energy was $5.499\;\rm{GeV}$, with a typical intensity of $7\;\rm{nA}$, incident
on a 5-cm liquid hydrogen target.  The signal from the 
scattered electron provided
the data-acquisition trigger.  The data-taking period
lasted for $42$ days and resulted in the collection of $\sim 4.3$ billion events.
After event reconstruction, $\sim 650$ million events remained with at least one
good electron candidate.

The scattered electron and associated hadrons were measured in the CLAS detector,
a large-acceptance magnetic spectrometer~\cite{clas} based on a six-coil toroidal
magnet with drift chambers providing
charged particle tracking, followed by a Cherenkov detector for electron identification,
and scintillators and an electromagnetic calorimeter for particle identification by time-of-flight and
energy deposition, respectively. 

The electron was identified by matching a negatively-charged
track in the drift chambers with signals in the Cherenkov counter
and in the electromagnetic calorimeter.
The identity of the positively-charged particle candidate was determined by
combining the flight time from the time-of-flight counters with
the momentum and track length from the drift chamber track to calculate 
the particle's velocity ($\beta$) and mass.

We analyzed events with a final state consisting of the scattered
electron plus one positively charged particle
(a $K^+$, $\pi^+$ or proton).  We measured the four-momenta of the scattered electron and 
charged hadron, and determined by missing-mass
that the undetected neutral particle was a $\Lambda$, a
neutron or a $\pi^0$, respectively.  

The scattered electron's and charged hadron's four momenta were
used to calculate the independent kinematic variables: \qsq, $W$,
\costh and $\phi$. 
Our kinematic coverages are $W=1.65-2.55$ GeV, $Q^2=1.6-4.6$ GeV$^2$ and 
the  full range of \costh and $\phi$.
We defined 720 bins in this four-dimensional space (see Table~\ref{binchoice}).
\begin{table}[bht]
\begin{center}
\begin{tabular}{|c|c|c|}
\multicolumn{3}{c}{} \\ \hline
{\bf Quantity} &{\bf No. Bin} &{\bf Bin Limits} \\ \hline\hline
$W$ (GeV)& 6 & 1.65, 1.75, 1.85, 1.95, 2.05, 2.25, 2.55   \\ \hline
$Q^2$ (GeV$^2$)& 2 & 1.6, 2.6, 4.6   \\ \hline
\costh & 5 & -1.0, -0.6, -0.2, 0.2, 0.6, 1.0\\ \hline
$\phi$ (deg) & 12 & -180., -150., ... 150., 180. \\ \hline
\end{tabular}\\
\caption[Kinematic binning]{\label{binchoice} Kinematic binning used in this analysis.}
\end{center}
\end{table}

For each event, the missing-mass recoiling from the scattered electron and identified
hadron was calculated and, by accumulating over all events, a missing-mass 
distribution was formed for each four-dimensional kinematic bin.  For the \ppi0 final state, 
an additional series of cuts was employed to remove radiative elastic-scattering events before our fits and mass cuts 
were applied.  We then fit each missing-mass distribution to a function 
consisting of a Gaussian peak for the signal and a smooth polynomial for the 
background.  We subtracted the background 
portion of the fit and counted the number
of events within a fixed missing-mass range to obtain the raw yield, using
the fit values for determination of the statistical uncertainty of the yield.

Corrections for finite acceptance and inefficiencies in track reconstruction,
particle identification and
missing-mass cuts were made.  A  Monte Carlo simulation, tuned to match the momentum resolution
of the detector, accounted for run-dependent inefficiencies due to malfunctioning
sub-system components.    

The acceptance-corrected yields were further corrected by a
two-body phase-space factor~\cite{frazer},
\begin{equation}
\Delta\rho_2 = {\bf |K_1|} / (16\pi^2 W)~,
\label{phasespace} 
\end{equation}
where ${\bf |K_1|}$, the momentum in the center-of-mass frame, and $W$ are evaluated at bin center.
We did not correct our data for radiative effects because explicit calculations 
showed that the
radiative correction factors for the \kpluslam~and \neutronpiplus~channels agreed within
$\pm 10\%$ for all bins~\cite{KPARK14}, which is smaller than the systematic uncertainty of the ratio, 
and showed no discernible kinematic dependence.
Some corrected yields for the \ppi0 channel were rejected for further analysis 
if the acceptance for the bin in question was lower than $2\%$.

The major sources and sizes of systematic uncertainties in the determination of the yields
are summarized in Table~\ref{corrections}, grouped by category. 
Overall, we assign a systematic uncertainty of
9\%, 18\% or 13\% to the $n\pi^+$, \ppi0 or \kpluslam corrected yields, respectively.

We then fit the $\phi$ distributions of the corrected yields in each bin of \qsq, $W$ 
and \costh to the form $A + B\cos2\phi + C\cos\phi$.
Some fits were rejected in the case of the
\ppi0 channel if there were fewer than 9 $\phi$ data points (of a nominal 12)
surviving the minimum acceptance cut.  This procedure resulted in
60 independent fitted values of the (A) terms for the \kpluslam~and
\neutronpiplus~channels, but only 48 for the
\ppi0 channel.
We divided the (A) terms for the different channels to form the 
cross-section ratios~\cite{binsize}.

\begin{table}
\begin{center}
\begin{tabular}{|cl|cr|cr|cr|} \hline
\multicolumn{2}{|c|}{\bf Procedure }   & \multicolumn{6}{c|}{\bf Systematic Uncertainty } \\
                                       &&\multicolumn{2}{c}{$n\pi^+$}  & \multicolumn{2}{c}{$p\pi^0$}  &\multicolumn{2}{c|}{$\Lambda K^+$} \\ \hline\hline

 \multicolumn{2}{|l|} {\bf Raw Yield Determination} & \multicolumn{2}{c|}{\bf 7\%}  & \multicolumn{2}{c|}{\bf 17\%}  & \multicolumn{2}{c|}{\bf 12\%} \\ \hline
  & Hadron PID cuts & & 3\%  & 10\%  & & \multicolumn{2}{c|}{ 11\%} \\ 
  & Missing-mass cuts   & & 3.5\% & 10.5\% & & \multicolumn{2}{c|}{2.5\%} \\ 
  & Background subtraction      & & 5\% & 6\% & & \multicolumn{2}{c|}{0.3\%} \\  \hline
 \multicolumn{2}{|l|} {\bf Efficiency Correction}  & \multicolumn{2}{c|}{\bf 6\%}  & \multicolumn{2}{c|}{\bf 5\%} & \multicolumn{2}{c|}{\bf 5\%} \\ \hline
  & Event generator dependence & & 1\% & 1\% & & \multicolumn{2}{c|}{0.7\%} \\  
  & Fiducial cuts       & & 2.5\% & 0.5\% & & \multicolumn{2}{c|}{0.2\%} \\ 
  & Trigger/Tracking eff.       & & 5\% & 5\% & & \multicolumn{2}{c|}{5\%} \\ \hline
 \multicolumn{2}{|l|} {\bf Phase Space Correction}& \multicolumn{2}{c|}{\bf 1.0\%}  & \multicolumn{2}{c|}{\bf 0.4\%}  & \multicolumn{2}{c|}{\bf 0.1\%} \\ \hline\hline
  \multicolumn{2}{|c|} {\bf Total Uncertainty}    & \multicolumn{2}{c|}{\bf 9\%}  & \multicolumn{2}{c|}{\bf 18\%} & \multicolumn{2}{c|}{\bf 13\%}     \\ \hline
\end{tabular}
\caption[Corrections to Yields]{\label{corrections} Sources and estimates of systematic 
uncertainties of the acceptance-corrected yields.}
\end{center}
\end{table}

\begin{figure}[!htb]
\begin{center}
  \includegraphics[angle=0,width=0.45\textwidth,height=12cm]{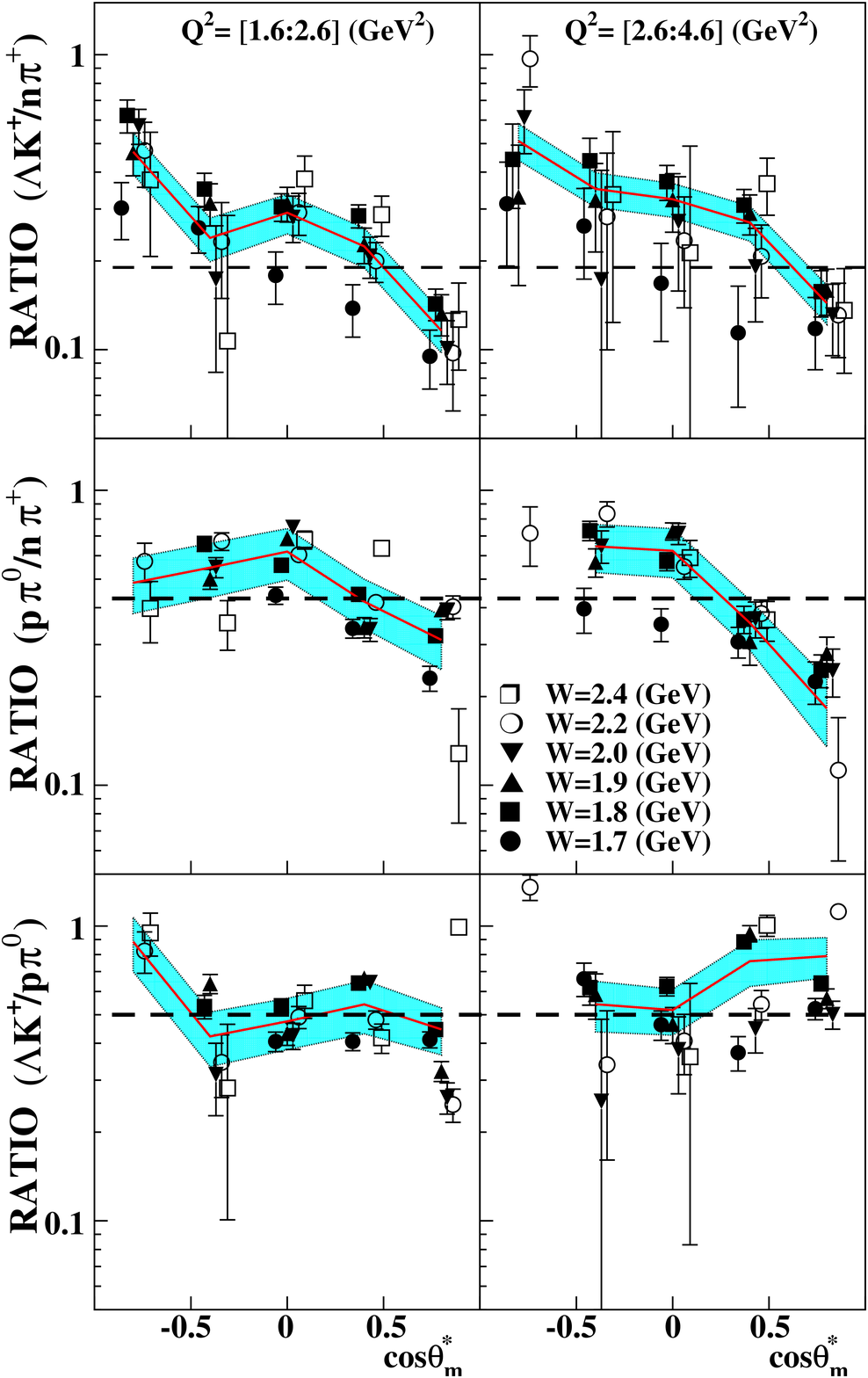}
  \caption[ratio_for_all]{\small{(Color online) Ratio of the three exclusive 
cross-sections, $\Lambda K^+$ to $n\pi^+$ (top), $p\pi^0$ to $n\pi^+$ (middle) and $\Lambda K^+$ 
to  $p\pi^0$ (bottom), 
for bins of $\langle Q^2 \rangle = 1.9$ GeV$^2$ (left) and $\langle Q^2 \rangle = 3.2$ GeV$^2$ (right), 
plotted versus \costh with different bins in $W$ shown as different symbols. 
The systematic uncertainty is indicated by the shaded band, centered on the solid (red) line which
connects the statistically-weighted average for each bin.  The flat dashed line represents the
overall statistical average for each ratio.
The data points are plotted slightly offset for clarity.}
\label{fig:ratio_for_all}  
  }        
\end{center}
\end{figure}
\begin{figure}[htb!]
\begin{center}
  \includegraphics[angle=0,width=0.45\textwidth, height=12cm]{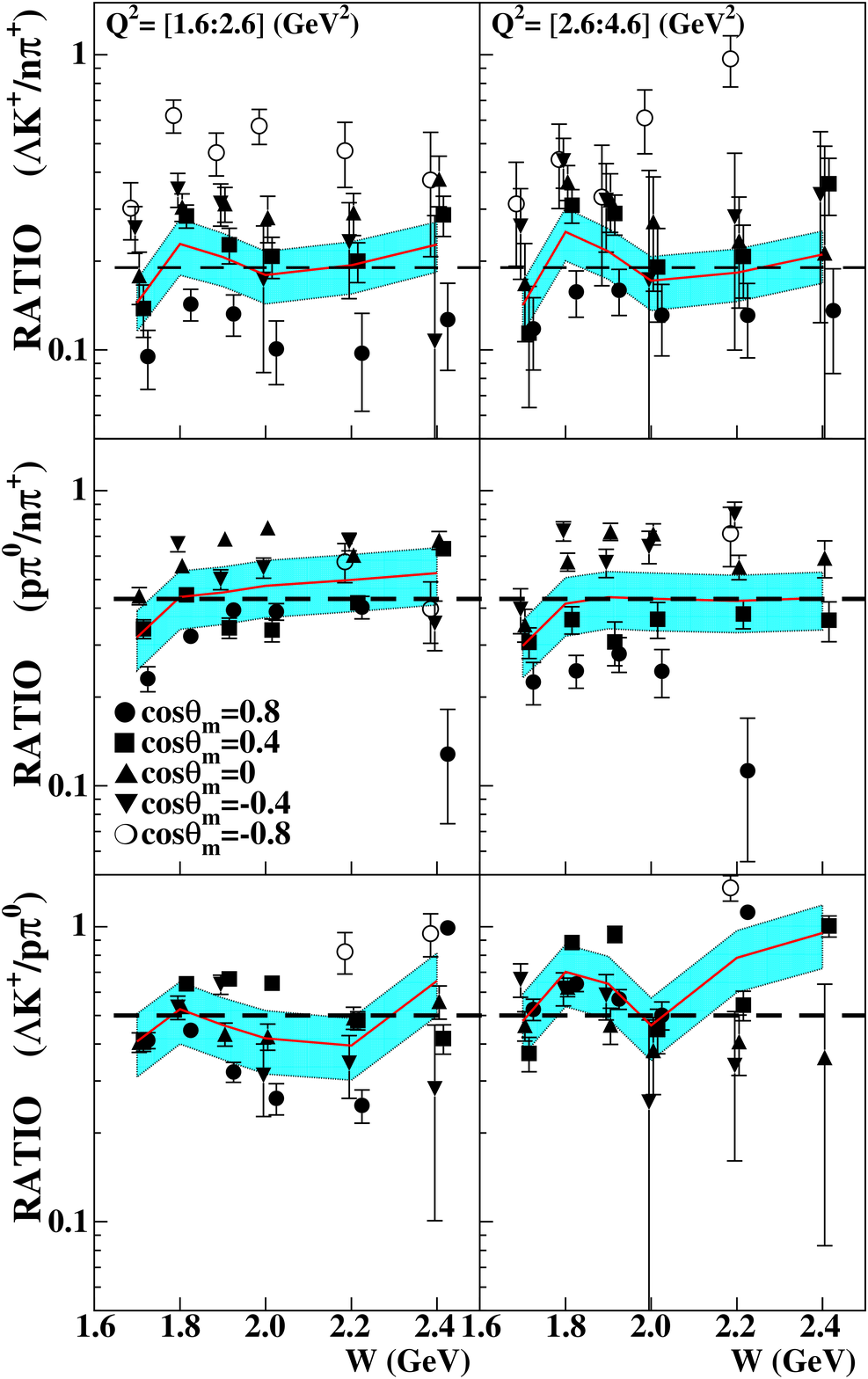}
  \caption[The ratio versus $W$]{\small{(Color online) Ratio of the three exclusive cross-sections,
      displayed as in Figure \ref{fig:ratio_for_all}, but plotted versus $W$ with different bins of \costh shown as different symbols.}
    \label{fig:ratio_for_all_w}
  }
\end{center}
\end{figure}

Figure~\ref{fig:ratio_for_all} shows the three ratios of corrected yields 
plotted versus \costh with the different symbols representing different $W$ bins.
The two columns show the $\langle Q^2 \rangle = 1.9$ GeV$^2$ bin (left) and 
the $\langle Q^2 \rangle = 3.2$ GeV$^2$ bin (right).
The shaded band is centered on the statistical average for each \costh bin with half-width
equal to the systematic uncertainty on the ratio.
Note that the three ratios are approximately the same for the two \qsq~bins
while there is a noticeable fall-off of the \kpluslam$/$\neutronpiplus~and 
\ppi0$/$\neutronpiplus~ratios with $\cos\theta^*_m$.
Figure~\ref{fig:ratio_for_all_w} shows the same ratios as in Figure~\ref{fig:ratio_for_all}, but plotted versus $W$.  Again the two columns are for the two bins in \qsq.  One can see that
the ratios are approximately independent of $W$.

For purposes of comparing with the single value of $\lambda_s$ used to characterize the ratio of strange
to non-strange hadronic production at high energy, we performed
a weighted average over all bins for each ratio of final states, indicated by
the flat dashed line.  We obtain the following average values for the ratios:
$\langle$\kpluslam$/$\neutronpiplus$\rangle  = 0.19 \pm 0.01 \pm 0.03$,
$\langle$\ppi0$/$\neutronpiplus$\rangle = 0.43 \pm 0.01 \pm 0.09$, and
$\langle$\kpluslam$/$\ppi0 $\rangle = 0.50 \pm 0.02 \pm 0.12$; the first
uncertainty is statistical and the second systematic.

%

We extracted the ratio of \qqbar creation probabilities from these
measured hadronic ratios 
using a simple factorization model in which a quark is knocked out of the proton
followed by a single \qqbar creation and appearance of the lightest baryon and
pseudo-scalar meson consistent with the quark flavor.
We ignored other processes such as vector-meson coupling to the 
virtual photon or $t$-channel exchange, which might be responsible for
the \costh dependence of our data.  Nevertheless, we hope that the results
from our simplified modelling are useful for comparison with strangeness suppression
results from semi-inclusive production experiments analysed under similar factorization
assumptions.
We note similarities of our model with that used by M.~M.~Kaskulov {\em et al.}~\cite{kaskulov} in fitting other data from
Jefferson Lab on electroproduction of $n\pi^+$ from the proton.

In our model, events are initiated by virtual photon absorption by a valence $u$-quark or 
$d$-quark in the ratio of the sums of squares of the quark charges (8:1). 
This is followed by a single \qqbar produced with probability ${\bf P}(q\overline q)$,  
resulting in a \qqbar state recoiling form a $qqq$ state.
Finally, the \qqbar state hadronizes into the lowest energy meson state and the 
$qqq$ state hadronizes into the lowest energy baryon state, in both cases 
with unit probability, resulting in three possible final states: \neutronpiplus, \ppi0 or \kpluslam.
Note also that we take into account that the $\pi^0$ is a $50:50$ mixture of \uubar and $d\overline d$.

Following this simple arithmetic, the hadronic production rates (${\Re }$) can be
written in terms of the \qqbar probabilities (${\bf P}(q\overline q)$) as such:

\noindent${\Re }(\Lambda K^+) \propto 8 \cdot {\bf P}(s\overline s)$,~~
${\Re }(n\pi^+) \propto 8 \cdot {\bf P}(d\overline d)$, and\\
${\Re }(p\pi^0) \propto 1/2\cdot\left( 8\cdot{\bf P}(u\overline u) + 1\cdot{\bf P}(d\overline d) \right)$.

We use the $\langle$\kpluslam$/$\neutronpiplus$\rangle$ ratio to solve for $s\overline s/d\overline d$ and the $\langle$\ppi0$/$\neutronpiplus$\rangle$ ratio to solve for $u\overline u/d\overline d$:
\begin{eqnarray}
s\overline s/d\overline d &=& \langle \Lambda K^+/n\pi^+\rangle = 0.19 \pm 0.01 \pm 0.03~,\nonumber\\
u\overline u/d\overline d &=& 2 ~ \left( \langle p\pi^0/n\pi^+ \rangle - 1/16 \right) = 0.74 \pm 0.02 \pm 0.18\nonumber
\end{eqnarray}
Finally, we use the $\langle$\kpluslam$/$\ppi0$\rangle$ ratio to determine an independent
measure of the \ssbar$/$\ddbar ratio.
We obtain

\noindent$s\overline s/$\ddbar = $1/2 ~ (u\overline u$/$d\overline d + 1/8)\langle$\kpluslam$/$\ppi0$\rangle$, yielding
\begin{eqnarray}
s\overline s/d\overline d = 0.28 \pm 0.01 \pm 0.07~ &&{\rm{assuming}}~ u\overline u/d\overline d = 1.0,
{\rm{or}}\nonumber\\
s\overline s/d\overline d = 0.22 \pm 0.01 \pm 0.07~ &&{\rm{assuming}}~ u\overline u/d\overline d = 0.74\nonumber\\
&&{\rm{(as~measured).}}\nonumber
\end{eqnarray}
The systematic uncertainty on a \qqbar ratio is simply that of the 
particle production ratio from which it is derived.  We do not include factors due to the angular dependence of the
ratios, nor do we attempt to quantify the systematic uncertainty of our hadronization model.
Table~\ref{results} summarizes the results of this extraction. 
\begin{table}[hb!]
\begin{center}
\begin{tabular}{|c|c|c|} \hline 
{\bf Ratio} & {$s\overline s/d\overline d$} & {$u\overline u/d\overline d$ }\\ \hline
{\bf {$\Lambda K^+$}/{$n\pi^+$}} & $0.19  \pm 0.03$ & -- \\ \hline
{\bf {$\Lambda K^+$}/{$p\pi^0$}  ``a''} & $0.22 \pm 0.07$ & -- \\ \hline
{\bf {$\Lambda K^+$}/{$p\pi^0$}  ``b''} & $0.28 \pm 0.07$ & -- \\ \hline
{\bf {$p\pi^0$}/{$n\pi^+$}} & -- &  $0.74 \pm 0.18$ \\ \hline
\end{tabular}
\caption[Table of Results]{\label{results} Values of ratio of \ssbar to \ddbar and \uubar to \ddbar shown according to the experimental ratios from which they are derived. The {\bf ``a''} and {\bf ``b''} cases for the $\Lambda K^+$/$p\pi^0$ data-set refer to the values of $u\overline u$/$d\overline d$ used in the extraction of the $s\overline s$/$d\overline d$ ratio: 0.74 for the {\bf ``a''} case and 1.0 for the {\bf ``b''} case. The uncertainties are the systematic uncertainties. }
\end{center}
\end{table}

We point out that our result of $0.74 \pm 0.18$ for the $u\overline u/d\overline d$ ratio
is different from the value of unity expected from isospin invariance arguments, 
as assumed, for example, in high-energy hadronization models.  However, we note
that our hadron-production environment is explicitly not isospin invariant
because the target is a proton, with two valence u-quarks and one valence d-quark.
We speculate that the isospin-dependence of our result for the $u\overline u/d\overline d$ 
ratio is related to the difference between the intrinsic $\overline u$ and $\overline d$ content 
of the proton as measured in Drell-Yan\cite{nusea} and semi-inclusive DIS 
experiments\cite{hermes}.
Although intriguing, unfortunately our measurement is not significantly different
from unity, especially when model uncertainties are included.

To summarize, our results show a sizable suppression of the $\Lambda K^+$ channel relative to the
$n\pi^+$ and \ppi0 channels from which we use a simple factorization model to estimate
a strangeness suppression factor ($s\overline s$/$d\overline d$)  
of $0.19 \pm 0.03$, $0.22 \pm 0.07$ or $0.28 \pm 0.07$,
depending on which data ratios we use and what we assume for the $u\overline u$$/$\ddbar ratio.
Interestingly, these values are similar to measurements of flavor suppression at high energies~\cite{suppressionresults},~\cite{zmasssuppression},~\cite{hera}.

These determinations of the flavor dependence of \qqbar creation are the first in the low-energy exclusive 
limit where the connection between the observed hadronic ratios and \qqbar production probabilities
is simple because only a single
\qqbar pair is created.  
However, further development of exclusive reaction theory is needed to reduce
the model dependence in the extraction of
the \qqbar creation probabilities from our data.
We conclude by noting that understanding \qqbar
production dynamics is an important part of understanding color confinement in QCD and
the fact that our values for strangeness suppression agree well with 
measurements done at much higher energy argues strongly for the 
universal nature of these dynamics.

We thank the staff of the Accelerator and Physics
Divisions at Jefferson Lab for making the experiment possible.
We also thank C.~Weiss
for many informative discussions.
This work was supported in part by the US Department of Energy,
the National Science Foundation, the Italian Istituto Nazionale
di Fisica Nucleare, the French American Cultural Exchange (FACE)
and Partner University Funds (PUF) programs, the French Centre
National de la Recherche Scientifique, the French Commissariat \`a l'Energie
Atomique, the United Kingdom's Science and Technology Facilities Council,
the Chilean Comisi\'on Nacional de Investigaci\'on Cient\'ifica y Tecnol\'ogica
(CONICYT), and the National Research Foundation of Korea. The Southeastern Universities Research Association
(SURA) operated the Thomas Jefferson National Accelerator Facility for the
US Department of Energy under Contract No.DE-AC05-84ER40150.


\begin{thebibliography}{0}
\bibitem{binsize} The bin sizes and virtual photon flux-factors
are the same for the three channels and thus divide out in the ratio.
\end{thebibliography}
\end{document}